\documentclass[aps,pre,twocolumn,showpacs,superscriptaddress]{revtex4-1}

\usepackage[final]{graphicx}
\usepackage{amsmath}
\usepackage{verbatim}
\usepackage{color}
\usepackage{ulem}

\usepackage{threeparttable}

\usepackage{multirow}
\usepackage[colorlinks,
linkcolor=blue,
citecolor=blue,
urlcolor=blue]{hyperref}

\begin{document}

\preprint{}

\title{Growth rate and gain of stimulated Brillouin scattering considering nonlinear Landau damping due to particle trapping}			

\author{Q. S. Feng} 
\affiliation{Institute of Applied Physics and Computational
	Mathematics, Beijing, 100094, China}

\author{L. H. Cao} 
\email{cao\_lihua@iapcm.ac.cn}
\affiliation{Institute of Applied Physics and Computational
	Mathematics, Beijing, 100094, China}
\affiliation{HEDPS, Center for
	Applied Physics and Technology, Peking University, Beijing 100871, China}
\affiliation{Collaborative Innovation Center of IFSA (CICIFSA) , Shanghai Jiao Tong University, Shanghai, 200240, China}

\author{Z. J. Liu} 
\affiliation{Institute of Applied Physics and Computational
	Mathematics, Beijing, 100094, China}
\affiliation{HEDPS, Center for
	Applied Physics and Technology, Peking University, Beijing 100871, China}

\author{L. Hao} 
\affiliation{Institute of Applied Physics and Computational
	Mathematics, Beijing, 100094, China}

\author{C. Y. Zheng} 
 \email{zheng\_chunyang@iapcm.ac.cn}
\affiliation{Institute of Applied Physics and Computational
	Mathematics, Beijing, 100094, China}
\affiliation{HEDPS, Center for
	Applied Physics and Technology, Peking University, Beijing 100871, China}
\affiliation{Collaborative Innovation Center of IFSA (CICIFSA) , Shanghai Jiao Tong University, Shanghai, 200240, China}

\author{C. Ning} 
\affiliation{Institute of Applied Physics and Computational
	Mathematics, Beijing, 100094, China}

%

\author{X. T. He} 
\affiliation{Institute of Applied Physics and Computational
	Mathematics, Beijing, 100094, China}
\affiliation{HEDPS, Center for
	Applied Physics and Technology, Peking University, Beijing 100871, China}
\affiliation{Collaborative Innovation Center of IFSA (CICIFSA) , Shanghai Jiao Tong University, Shanghai, 200240, China}


\date{\today}

\begin{abstract}
Growth rate and gain of SBS considering the reduced Landau damping due to particle trapping has been proposed to predict the growth and average level of SBS reflectivity. Due to particle trapping, the reduced Landau damping has been taken used of to calculate the gain of SBS, which will make the simulation data of SBS average reflectivity be consistent to the Tang model better. This work will solve the pending questions in laser-plasma interaction and have wide applications in parametric instabilities.

\end{abstract}


\maketitle


Backward stimulated Brillouin scattering (SBS) is a three-wave interaction process where an incident electromagnetic wave (EMW) decays into a backscattered EMW and a forward propagating ion-acoustic wave (IAW), which will lead to a great energy loss of the incident laser and is detrimental in inertial confinement fusion (ICF) \cite{He_2016POP,Glenzer_2010Science,Glenzer_2007Nature}. In the indirect-drive ICF \cite{Glenzer_2010Science,Glenzer_2007Nature} or the hybrid-drive ignition \cite{He_2016POP,Lan_2016MRE,Huo_2016MRE,Huo_2016PRL}, CH was chosen as the standard ablator material for ICF ignition capsules due to its low atomic number, high density, and a host of manufacturing considerations. Thus, the inside of hohlraum will be filled with low-Z plasmas, such as H or CH plasmas from the initial filled material or from the ablated material off the capsule. 

It is one of the key issues for the success of laser fusion to control backward SBS. Many mechanisms for the saturation of SBS have been proposed, including increasing linear Landau damping by kinetic ion heating \cite{Rambo_1997PRL, Pawley_1982PRL}, frequency detuning due to particle trapping \cite{Froula_2002PRL}, coupling with higher harmonics \cite{Bruce_1997POP, Rozmus_1992POP}, the creation of cavitons in plasmas \cite{ Weber_2005PRL, Weber_2005POP,Liu_2009POP1} and so on. One of the pending questions in laser-plasma interaction is the discrepancy between the theoretically predicted reflectivity and the experimentally observed reflectivity. \cite{Berger_2015PRE}
SBS reflectivity is directly proportional to the IAW amplitude, and SBS driven IAW behaves in a nonlinear way, such as particle trapping and harmonic generation. Particle trapping may produce a non-Maxwellian distribution, reducing Landau damping of IAW \cite{Neil_1965POF} and potentially eliminating the higher IAW damping of the multi-ion species plasmas. Therefore, it is not applicable to predicting SBS reflectivity in experiment by the traditional linear growth rate and gain of SBS taking use of only the linear Landau damping, where the particle trapping is not considered and the Maxwell distribution is assumed. On the other hand, harmonic generation will dissipate the energy of fundamental IAW mode to harmonics, which will induce an efficient damping to decrease the growth rate of SBS. \cite{Rozmus_1992POF,Cohen_1997POP} However, when the harmonic amplitude is large, the pump depletion will be obvious, thus the SBS will be saturated mainly due to pump depletion.

 In this Rapid Communication, we report the first demonstration that the growth rate and gain of SBS considering the nonlinear Landau damping of IAW due to particle trapping will be consistent to the simulation results better. This model considering the nonlinear Landau damping will give a good explanation of why the reflectivity in experiment or simulation was higher than the prediction of Tang model \cite{Tang_1966JAP}, where Berger et al. \cite{Berger_2015PRE} modeled the weakly damped data with a  \textquotedblleft speckle-enhanced" $G=2G(I_0)$, twice the calculated gain. This Rapid Communication will show that the Landau damping will decrease with time due to particle trapping, thus the simulation data will be modeled well with gain of SBS considering reduced Landau damping due to particle trapping.

\begin{figure}[!tp]
	\includegraphics[width=0.8\columnwidth]{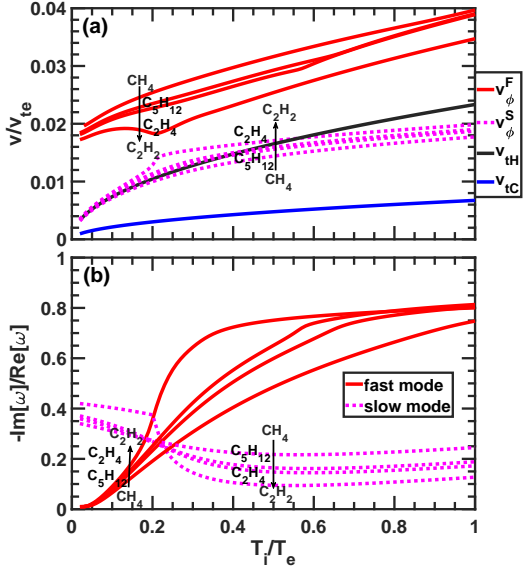}
	
	\caption{\label{Fig:Dispersion}(Color online) (a) The phase velocity and (b) the Landau damping $-Im[\omega]/Re[\omega]$ of the fast mode and the slow mode in CH$_4$, C$_5$H$_{12}$, C$_2$H$_4$ and C$_2$H$_2$ plasmas. The arrow refers to the line from CH$_4$, C$_5$H$_{12}$, C$_2$H$_4$ to C$_2$H$_2$ in order. The conditions are: $T_e=5keV$, $n_e=0.3n_c$ and $k_{A}\lambda_{De}=0.3$.}
\end{figure}

The IAW frequency is much lower than that of SBS scattering light. Thus,
the wave number of IAW excited by backward SBS is $
\label{Eq:k_A}
k_A\lambda_{De}\simeq 2k_0=2\frac{v_{te}}{c}\sqrt{n_c/n_e-1},
$
where $k_0$ is the wave number of pump light, $v_{te}=\sqrt{T_e/m_e}$ is the electron thermal velocity, $n_e, T_e, m_e$ are the density, temperature and mass of the electron.
Assuming fully ionized, neutral, unmagnetized plasmas, the linear dispersion relation of the ion acoustic wave in multi-ion species plasmas is given by \cite{Williams_1995POP,Feng_2016POP,Feng_2016PRE}
\begin{equation}
\label{Eq:Dispersion}
\epsilon(\omega,k=k_A)=1+\sum_j \chi_{j}=0,
\end{equation}
where $\chi_{j}$ is the susceptibility of particle $j$ ($j=e, H, C$). And $\omega=Re(\omega)+i*Im(\omega)$ is complex frequency. $Re(\omega)$ and $Im(\omega)$ are the frequency and Landau damping of IAW.
Under the condition of $T_e=5keV$, $n_e=0.3n_c$, one can obtain the wave number of the IAW $k_A\lambda_{De}=0.3$. By solving Eq. (\ref{Eq:Dispersion}), the phase velocity $v_{\phi}=Re(\omega)/k$ and Landau damping $-Im[\omega]/Re[\omega]$ of the fast mode and the slow mode in CH$_4$, C$_5$H$_{12}$, C$_2$H$_4$ and C$_2$H$_2$ are given as shown in Fig. \ref{Fig:Dispersion}.
There exist two groups of modes called \textquotedblleft fast mode" and \textquotedblleft slow mode" in multi-ion species plasmas as shown in Fig. \ref{Fig:Dispersion}. Here, the fast mode refers to the mode with phase velocity larger than the thermal velocity of each species, and the slow mode refers to the mode with phase velocity close to the thermal velocity of light species ions as shown in Fig. \ref{Fig:Dispersion}(a). When $T_i/T_e\lesssim 0.2$, the Landau damping of the fast mode is lower than that of the slow mode, thus the fast mode is the dominated mode. When $T_i/T_e\gtrsim0.2$, the Landau damping of the slow mode is lower than that of the fast mode, thus the slow mode is the dominated mode. In this Rapid Communication,  $T_i/T_e=0.5$ is chosen as the typical parameter since $T_i$ approaches $T_e/2$ at the peak laser power in ICF ignition experiments \cite{Meezan_2010POP}. With the ratio of C to H increasing, i.e., from CH$_4$, C$_5$H$_{12}$, C$_2$H$_4$ to C$_2$H$_2$, the Landau damping of the slow mode will decrease.

\begin{figure}[!tp]
	\includegraphics[width=0.8\columnwidth]{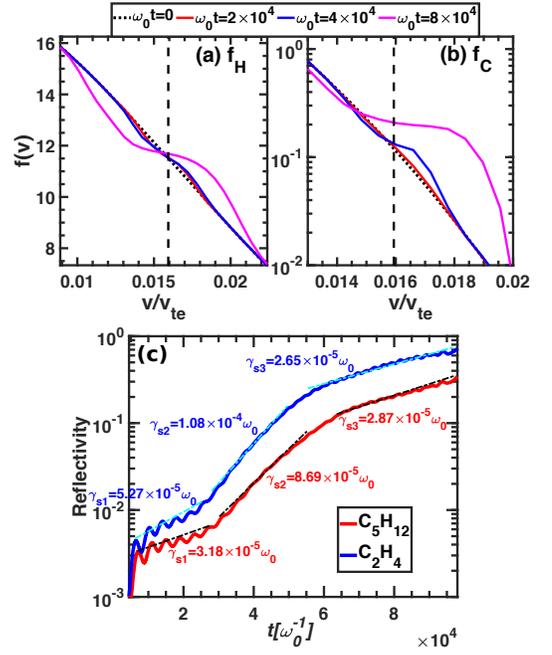}
	
	\caption{\label{Fig:Growth_R}(Color online) 
	The velocity distribution functions averaged in the total spatial scale of (a) H ions and (b) C ions in C$_2$H$_4$ plasmas at different time points. (b) The growth rate of SBS reflectivity in C$_5$H$_{12}$ and C$_2$H$_4$ plasmas. The conditions are: $T_e=5keV$, $T_i=0.5T_e$, $n_e=0.3n_c$ and $k_{A}\lambda_{De}=0.3$ in C$_2$H plasmas.}
\end{figure}

A one dimension Vlasov-Maxwell code \cite{Liu_2009POP} is used to research the SBS nonlinear growth rate and gain when the trapping and harmonic generation processes are considered. 
The electron temperature is $T_e=5keV$ and electron density is $n_e=0.3n_c$, where $n_c$ is the critical density for the incident laser. The electron density is taken to be higher than $0.25n_c$, thus the stimulated Raman scattering \cite{Feng_2018POP} and two-plasmon decay instability \cite{Xiao_2015POP,Xiao_2016POP} are excluded. The CH$_4$, C$_5$H$_{12}$, C$_2$H$_4$, C$_2$H$_2$ plasmas are taken as typical examples since they are common in ICF \cite{He_2016POP,Glenzer_2007Nature}. The ion temperature is $T_i=0.5T_e$ and the slow mode will be excited and dominate in SBS. The linearly polarized pump laser intensity is $I_0=3\times10^{15}W/cm^2$ and the wavelength is $\lambda_0=0.351\mu m$. And the seed light from the right boundary is with the intensity of $I_s=10^{-4}I_0=3\times10^{11}W/cm^2$ and the matching frequency. The spatial scale is [0, $L_x$] discretized with $N_x=5000$ spatial grid points and spatial step $dx=0.2c/\omega_0$. And the spatial length is $L_x=1000c/\omega_0\simeq160\lambda_0$ with $2\times5\%L_x$ vacuum layers and $2\times5\%L_x$ collision layers in the two sides of plasmas boundaries. The plasmas located at the center with density scale length $L=0.8L_x$ are collisionless. The boundary condition of incident laser is open. The strong collision damping layers are added into the two sides of the plasmas boundaries ($2\times5\%L_x$) to damp the electrostatic waves such as IAWs at the boundaries and decrease the effect of sheath field. The electron velocity scale $[-0.8c, 0.8c]$ and the ion (C and H) velocity scale $[-0.03c, 0.03c]$ are discretized with $2N_v+1$ ($N_v=512$) grid points. The total simulation time is $t_{end}=1\times10^5\omega_0^{-1}$ discretized with $N_t=5\times10^5$ and time step $dt=0.2\omega_0^{-1}$.

\begin{figure}[!tp]
	\includegraphics[width=1\columnwidth]{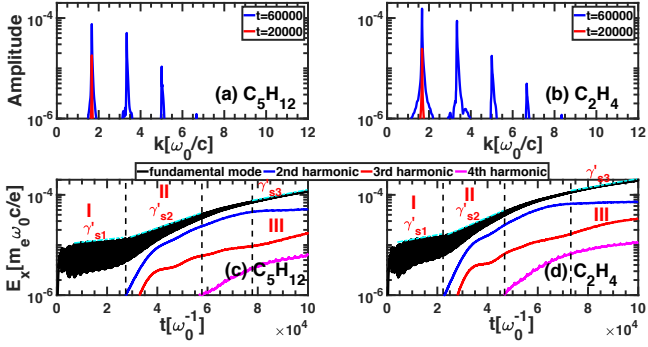}
	
	\caption{\label{Fig:Spectra_w_Ex}(Color online) The wave number spectra of $E_x$ in (a) C$_5$H$_{12}$ and (b) C$_2$H$_4$ plasmas at the time of $\omega_0t=2\times10^4$ and $\omega_0t=6\times10^4$.
	The time evolution of fundamental mode and harmonics in (c) C$_5$H$_{12}$ and (d) C$_2$H$_4$ plasmas.
		 The condition is as the same as Fig. \ref{Fig:Growth_R}.}
\end{figure}

  \begin{figure}[!tp]
  	\includegraphics[width=0.8\columnwidth]{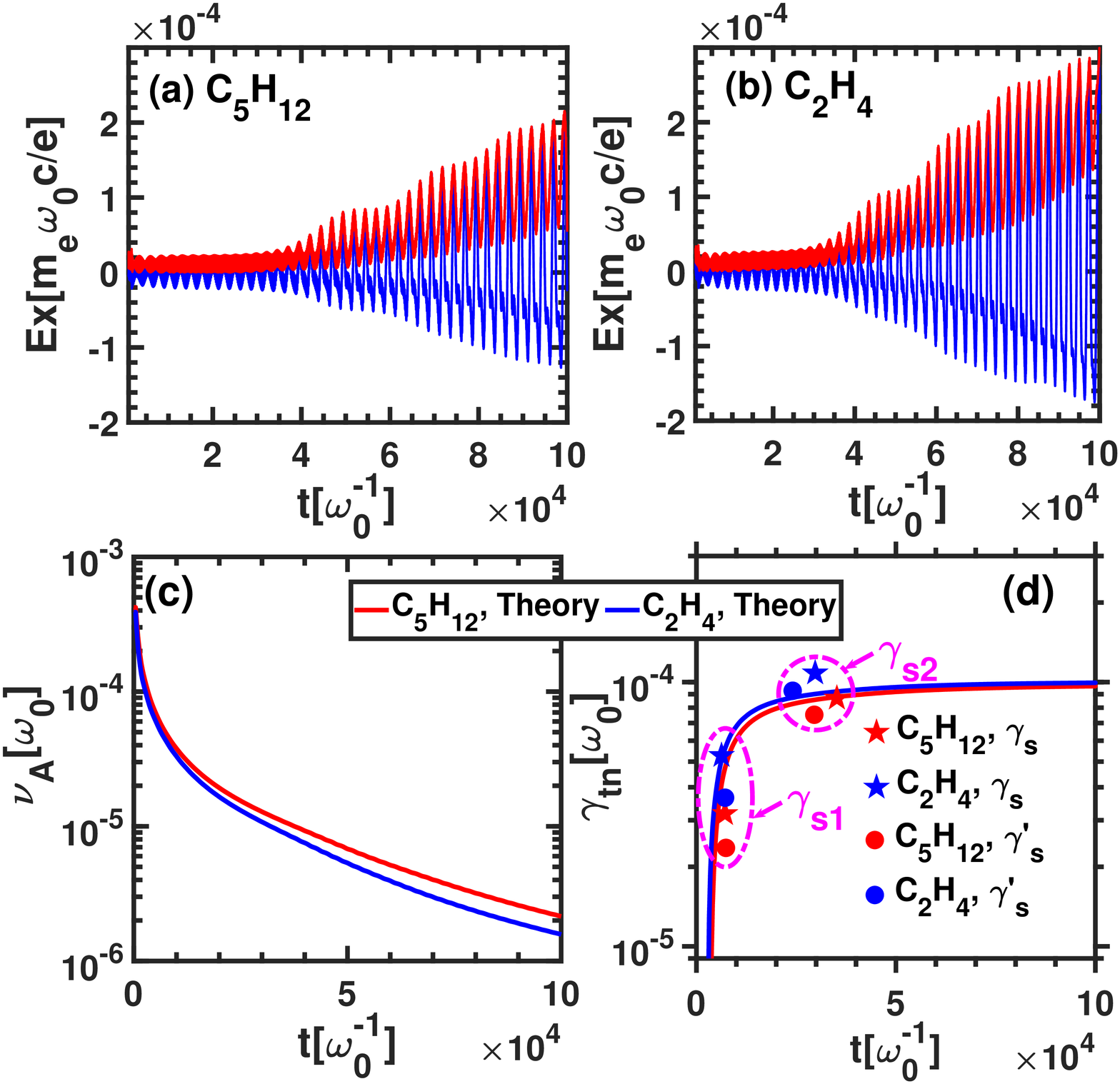}
  	\caption{\label{Fig:gamma_tn}(Color online) The time evolution and envelop of electrostatic field $E_x$ in (a) C$_5$H$_{12}$ and (b) C$_2$H$_4$ plasmas. The time evolution of (c) Landau damping of the slow mode and (d) growth rate considering the nonlinear Landau damping in C$_5$H$_{12}$ and C$_2$H$_4$ plasmas.
  	Where $\gamma_{s}$ are the simulation data calculated from Fig. \ref{Fig:Growth_R}(c) and $\gamma'_{s}$ are the simulation data taken from Figs. \ref{Fig:Spectra_w_Ex}(c) and \ref{Fig:Spectra_w_Ex}(d). The start time points of growth rates $\gamma_{s1}$ and $\gamma_{s2}$ are approximately chosen as the time of simulation data. }
  \end{figure}

 Figure \ref{Fig:Growth_R} gives the distribution functions in C$_2$H$_4$ plasmas at different time points and the reflectivity of SBS in C$_2$H$_4$ and C$_5$H$_{12}$ plasmas. Taking C$_2$H$_4$ plasmas as an example, the absolute value of slope of distribution function around the phase velocity will decrease with time due to particle trapping as shown in Figs. \ref{Fig:Growth_R}(a) and \ref{Fig:Growth_R}(b). Since the Landau damping is proportional to the slope of distribution function at the phase velocity, the Landau damping will decrease through particle trapping with time increasing. The theoretical growth rate of the SBS scattered light in homogeneous plasmas is given by \cite{LiuCS_1974POF,DuBois_1974PRL}
 \begin{equation}
 \label{Eq:growth rate}
 \gamma_t=[\frac{2\gamma_{0B}}{\sqrt{|v_{gs}|v_{gA}}}-(\frac{\nu_s}{|v_{gs}|}+\frac{\nu_A}{v_{gA}})]\frac{|v_{gs}|v_{gA}}{|v_{gs}|+v_{gA}},
 \end{equation}
 where
 $
 \gamma_{0B}=\frac{1}{4}\sqrt{\frac{n_e}{n_c}}\frac{v_0}{v_{te}}\sqrt{\omega_0\omega_A}
 $ 
 is the maximum temporal growth rate of SBS \cite{Berger_2015PRE,Lindl_2004POP},  $v_0=eA_0/m_ec$ is the electron quiver velocity. $v_{gs}$, $v_{gA}$ are the group velocity of SBS scattering light and IAW.
The damping rate of the backscattered light $\nu_s$ is negligible since it is much lower than the IAW Landau damping $\nu_A$, i.e., $\nu_s\simeq0$. The simplified expression of the nonlinear IAW Landau damping is given by \cite{Yampolsky_2009POP1,Yampolsky_2009POP2,Wangq_2018POP}
\begin{equation}
\label{Eq:nonlinear Landau damping}
\nu_A(t)=\frac{\nu_A(0)}{1+\frac{3\pi^2}{128}\int\limits_{0}^{t}\omega_B(t)dt},
\end{equation}
where $\nu_A(0)$ is the linear Landau damping of IAW, $\omega_B(t)=\sqrt{eE_x(t)k_A/m_e}$ is the bounce frequency of electrons.
 As shown in Fig. \ref{Fig:Growth_R}(c), the linear growth of SBS reflectivity includes about three process: in stage I, the SBS grows with a low growth rate, the growth rate of SBS in C$_5$H$_{12}$ plasmas is $\gamma_{s1}=3.18\times10^{-5}\omega_0$ and that in C$_2$H$_4$ plasmas is $\gamma_{s1}=5.27\times10^{-5}\omega_0$ from the start time $t_{s1}\simeq7.5\times10^{3}\omega_0^{-1}$. The growth rate of SBS reflectivity in C$_2$H$_4$ plasmas is higher than that in C$_5$H$_{12}$ plasmas is due to the lower Landau damping in C$_2$H$_4$ plasmas. In stage II, the SBS will increase with a very large growth rate from $t_{s2}\simeq3\times10^4\omega_0^{-1}$. In C$_5$H$_{12}$ plasmas, $\gamma_{s2}=8.69\times10^{-5}\omega_0$ and in C$_2$H$_4$ plasmas, $\gamma_{s2}=1.08\times10^{-4}\omega_0^{-1}$. In stage III, the SBS growth rate will decrease and the SBS reflectivity will saturate after stage III. Stage I and stage II can be explained by the decrease of Landau damping of IAW due to particle trapping, and stage III is as a result of pump depletion and harmonic generation, which will be discussed below in detail.
 
 As shown in Figs. \ref{Fig:Spectra_w_Ex}(a) and \ref{Fig:Spectra_w_Ex}(b), at the early time such as $\omega_0t=2\times10^4$, only the fundamental mode appears  and the harmonics will appear in the later time such as $\omega_0t=6\times10^4$. From Figs. \ref{Fig:Spectra_w_Ex}(c) and \ref{Fig:Spectra_w_Ex}(d), there exist three processes of linear growth of the fundamental mode. In stage I, only the fundamental mode exists and no harmonics develop. Therefore, only the particle trapping plays a role as the nonlinear effect on the growth rate of fundamental mode. In stage II, the second and the third harmonics will increase with time, however, the amplitudes of harmonics are very low. Thus, the harmonic effect on the growth rate of $|E_x|$ is not obvious. The main nonlinear effect on growth rate of $|E_x|$ is due to particle trapping. After stage II, the nonlinear Landau damping of IAW due to particle trapping will decrease to nearly zero, as a result, the particle trapping can no longer determine the growth rate of $|E_x|$. Thus, with the harmonic increasing, the efficient damping of IAW from harmonic generation will increase and the growth rate of $|E_x|$ will be lower and lower. In stage III, the harmonics especially the second harmonic will saturate and no longer increase, thus the efficient damping of IAW due to harmonic generation will no longer increase and the growth rate of $|E_x|$ will keep constant with a low value $\gamma'_{s3}$. At the same time, SBS reflectivity reaches to a large level, thus the pump depletion will make the SBS saturation. 
  
  Figure \ref{Fig:gamma_tn} gives the time evolution of $E_x$, nonlinear Landau damping of slow mode $\nu_A$ and growth rate considering nonlinear Landau damping $\gamma_{tn}$ in C$_5$H$_{12}$ and C$_2$H$_4$ plasmas. From Eq. (\ref{Eq:nonlinear Landau damping}), $\omega_B(t)=\sqrt{eE_x(t)k_A/m_e}$ is related to the amplitude of $E_x(t)$, thus, nonlinear Landau damping is a function of time. The nonlinear Landau damping $\nu_A$ is calculated from Eq. (\ref{Eq:nonlinear Landau damping}) by integral of envelop of $E_x(t)$, and growth rate $\gamma_{tn}$ is calculated from Eq. (\ref{Eq:growth rate}) by considering nonlinear Landau damping. Comparing the growth rates of SBS reflectivity $\gamma_{s1}$, $\gamma_{s2}$ calculated from Fig. \ref{Fig:Growth_R}(c) with the theoretical growth rate considering nonlinear Landau damping due to particle trapping, one can see that the simulation results are close to the theoretical curve. In the same way, the growth rates of $|E_x|^2$ calculated from Fig. \ref{Fig:Spectra_w_Ex}(c) and \ref{Fig:Spectra_w_Ex}(d) are also consistent to the theoretical curve. Where $\gamma'_s$ labelled in Fig. \ref{Fig:gamma_tn}(d) is the growth rate of energy of electrostatic field $|E_x|^2$, i.e., $\gamma'_s=[ln(|E_x^{end}|^2)-ln(|E_x^{start}|^2)]/(t^{end}-t^{start})$ is twice of $\gamma'_s$ labelled in Figs. \ref{Fig:Spectra_w_Ex}(c) and \ref{Fig:Spectra_w_Ex}(d). These results illustrate that the growth rates in stage I and stage II will be mainly affected by particle trapping. However, in stage III, the growth rate will be lower than the theoretical curve considering nonlinear Landau damping only due to particle trapping (not shown in Fig. \ref{Fig:gamma_tn}). After stage II, particle trapping will make the Landau damping to be nearly zero, the harmonic generation will induce an efficient damping and the pump depletion will occur, which will make the growth rate lower as explained in Fig. \ref{Fig:Spectra_w_Ex}.

\begin{figure}[!tp]
	\includegraphics[width=0.8\columnwidth]{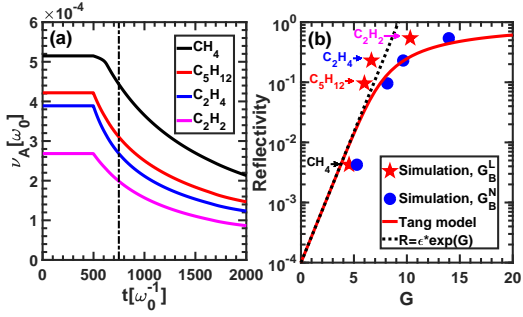}
	\caption{\label{Fig:R_G}(Color online) The early time evolution of (a) Landau damping of the slow IAW mode in different species plasmas. (b) The relation between the SBS reflectivity and SBS gain in different species plasmas. Where the Vlasov simulation data take the gain $G_B^L$ by considering linear Landau damping (pentagram points) and the gain $G_B^N$ considering nonlinear Landau damping at the time $\omega_0t=750$ (circle points).}
\end{figure}

Furtherly, the gain is calculated by considering the linear Landau damping of IAW and nonlinear Landau damping of IAW due to particle trapping as shown in Fig. \ref{Fig:R_G}. Since the pump light meets with the seed light at $t_0=500\omega_0^{-1}$, the Landau damping before $t_0=500\omega_0^{-1}$ can be thought as the linear Landau damping $\nu_A^L$ due to very low amplitude of IAW. Among the spatial scale $L_x=1000c/\omega_0$, the average time point of pump light interacting with the seed light is at $t=750\omega_0^{-1}$. Therefore, $t=750\omega_0^{-1}$ can be taken as a time point to calculate the nonlinear Landau damping and gain considering nonlinear Landau damping. From Eq. (\ref{Eq:nonlinear Landau damping}), $\nu_A(t=750\omega_0^{-1})$ in different species plasmas can be obtained as shown in Fig. \ref{Fig:R_G}(a). The gain of SBS by fluid theory is given by
\begin{equation}
\label{Eq:G_B}
G_B=2\frac{\gamma_{0B}^2}{\nu_Av_{gs}}L.
\end{equation}
Under the strong damping condition $\nu_A/\gamma_{0B}*\sqrt{v_{gs}/v_{gA}}\gg1$ \cite{Forslund_1975POF}, one can get the Tang model \cite{Tang_1966JAP}:
\begin{equation}
R(1-R)=\varepsilon\{\text{exp}[G(1-R)]-R\},
\end{equation}
where $R$ is the reflectivity of SBS at the left boundary, and $\varepsilon$ is seed light at the right boundary. Pump depletion has been considered in Tang model. If $R\ll1$, the Tang model can be approximate to the seed amplification equation:
\begin{equation}
\label{Eq:seed amplification}
R=\varepsilon\cdot\text{exp}(G).
\end{equation}
The gain of SBS can be obtained by considering the linear Landau damping of IAW $\nu_A^L$ and nonlinear Landau damping of IAW at $\omega_0t=750$ $\nu_A(t=750\omega_0^{-1})$, which are labelled as $G_B^L$ and $G_B^N$ respectively as shown in Fig. \ref{Fig:R_G}(b). When only the linear Landau damping of IAW is considered, the points by the Vlasov simulation are not consistent to the theoretical curve, especially when the gain is large, such as in C$_5$H$_{12}$, C$_2$H$_4$, C$_2$H$_2$ plasmas. However, when the nonlinear Landau damping is considered, the SBS gain from simulation will be revised to G$_B^N(t=750\omega_0^{-1})$. And the points from Vlasov simulation are consistent to the theoretical curve of Tang model by considering nonlinear Landau damping due to particle trapping. These results may give a good explanation of why the SBS reflectivity in experiments is always larger than the linear theory prediction.

In conclusions, growth rate and gain of SBS are proposed by considering the nonlinear Landau damping of IAW due to particle trapping. The simulation results are consistent to the theoretical analyses. Due to particle trapping, the Landau damping will decrease with time. The early SBS growth rate is mainly affected by the particle trapping, while in the later time, the harmonic generation and pump depletion will play a main role in reducing the SBS growth rate. When nonlinear Landau damping of IAW due to particle trapping is considered, the modified growth rate and gain of SBS  can predict the SBS average reflectivity more accurately, which will have a wide application in the field of parametric instability in ICF experiment.

We would like to acknowledge useful discussions with C. Z. Xiao, Q. Wang, W. D. Zheng and S. Y. Zou. This research was supported by National Postdoctoral Program for Innovative Talents (No. BX20180055), the China Postdoctoral Science Foundation (Grant No. 2018M641274), the National Natural Science Foundation of China (Grant Nos. 11875091, 11875093, 11675025 and 11575035), and Science Challenge Project, No. TZ2016005.

\bibliography{SBS_multi_species.bib}

\end{document}